\documentclass[10pt,twocolumn]{article}
\usepackage[a4paper,top=0.65in,bottom=0.68in,left=0.68in,right=0.68in]{geometry}
\usepackage[T1]{fontenc}
\usepackage[utf8]{inputenc}
\usepackage{lmodern}
\usepackage{microtype}
\usepackage{booktabs}
\usepackage{tabularx}
\usepackage{array}
\usepackage{float}
\usepackage{enumitem}
\usepackage{hyperref}
\usepackage{xurl}
\setlist{nosep,leftmargin=*}
\hypersetup{colorlinks=true,linkcolor=black,citecolor=black,urlcolor=blue}

\title{\vspace{-1.1em}\textbf{A Black Box for Agentic Processes:}\\
\large Blockchain-Anchored Evidence for AI Agent Communication, Human Oversight, and GRC Audits\vspace{-0.6em}}
\author{Arslan Br\"omme\\[-0.15em]
{\small\textit{CISSP, CISM, CISA, CAISE}}\\
\small Independent Researcher\\
\small \href{mailto:arslan.broemme@aviomatik.de}{arslan.broemme@aviomatik.de}\\
\small Draft v0.9.1.5 -- 3 September 2026\\[0.55em]
\begingroup
\setlength{\fboxsep}{5pt}%
\setlength{\fboxrule}{0.4pt}%
\fbox{\begin{minipage}{0.82\textwidth}
\centering\small\textit{Preprint / working paper}. This version is a work in progress and may be updated. Comments are welcome.
\end{minipage}}
\endgroup}
\date{}

\begin{document}
\maketitle
\vspace{-1.3em}

\begin{abstract}
Autonomous AI agents increasingly communicate with other agents, invoke tools, exchange intermediate results, and request human approvals. These workflows create a new auditability problem: organizations must reconstruct what happened, when it happened, which agent or human was involved, which control or policy applied, and whether records were modified afterwards. Motivated by the 2026 OpenAI/Hugging Face incident, this position and architecture paper proposes a product- and vendor-neutral black-box architecture for agentic processes. The architecture creates blockchain-anchored cryptographic commitments for selected agent communications, human-in-the-loop approvals, tool calls, and process artifacts without placing sensitive content on-chain. We define an evidence model that distinguishes temporal anchoring and artifact integrity from event ordering, capture authenticity, authorized anchoring, and causal traceability. The latter properties require additional architectural controls. We then discuss practical use for Governance, Risk, and Compliance (GRC), including compliance testing, risk-based evidence selection, monitoring evidence streams, incident reconstruction, and regulatory reporting readiness under the EU AI Act, NIS2, and the Cyber Resilience Act (CRA). This position and architecture paper does not present an empirical performance or security evaluation. The approach does not prevent agent misbehavior or prove semantic truth. Rather, it strengthens the evidentiary basis for later verification of critical process traces.
\end{abstract}

\section{Introduction}
AI systems are moving from passive assistants toward agentic architectures that can plan, communicate, call tools, manipulate files, create tickets, modify code, and interact with humans or other agents. This shift creates an evidence problem. It is no longer sufficient to ask what a human decided or what an application logged. In agentic workflows, organizations may need to establish which agent communicated with which other agent, which information was exchanged, which tool was invoked, what human approval was obtained, which policy applied, and whether the record of those steps was later modified.

The 2026 OpenAI/Hugging Face incident demonstrates why this question is practical rather than theoretical. OpenAI reported that models operating during internal cybersecurity evaluations circumvented isolation controls, communicated through unauthorized channels, exploited shared infrastructure, gained internet access, and accessed third-party systems \cite{openai2026hf}. An independent investigation by METR and Redwood Research reported that roughly 1,200 agents used an unsanctioned message board to exchange more than 70,000 messages and files. Approximately 700 participated in the Hugging Face attack. The investigators documented attempts to spoof, edit, or delete transcript evidence and reported limited successful demonstrations of tool-call spoofing \cite{metrredwood2026hf}. These findings motivate a broader requirement: evidence about agentic actions should not depend exclusively on mutable records inside the environment the agents can influence.

We therefore propose a \emph{black box for agentic processes}. ``Black box'' is used here in the event-recorder sense: an externalized, tamper-evident record of selected critical process states that supports reconstruction after an incident or during an audit. The guiding question is how an external evidence layer can provide integrity, temporal anchoring, and declared ordering for selected agentic events without exposing sensitive content. The approach is intentionally product- and vendor-independent. It describes an architectural pattern, not a specific commercial implementation or empirical evaluation. The contribution is not a new containment mechanism. Instead, it is an evidence architecture that applies externally verifiable cryptographic commitments to selected events in agentic workflows.

\section{Background, Related Work, and Threat Model}
Cryptographic hashing enables a verifier to recompute the digest of an artifact and compare it with a previously committed reference. If the digests match, the verifier gains evidence that the retained artifact is the same byte sequence as the artifact committed earlier. When the commitment is anchored in an append-only or blockchain-based system, later changes to the artifact can be detected without revealing the artifact itself.

This evidentiary claim is intentionally limited. Hash anchoring can support an \emph{existence and integrity proof}: a particular artifact existed in a particular form at or before the commitment time and has not changed without detection. It does not automatically prove authorship, authorization, truthfulness, or correctness. This distinction is useful for agentic evidence: integrity and temporal existence can be verified while identity, authorization, and accountability remain handled by separate enterprise controls.

Agentic workflows produce evidence-bearing artifacts such as agent-to-agent messages, human approval requests, tool-call parameters, results, intermediate files, tickets, policy decisions, and system-state snapshots. Many are sensitive and should remain off-chain. Their cryptographic commitments, however, can be anchored externally.

\paragraph{Related work.}
The proposed architecture builds on digital timestamping, tamper-evident logs, and blockchain anchoring. Haber and Stornetta introduced practical mechanisms for time-stamping digital documents to make back-dating or forward-dating difficult \cite{haber1991timestamping}. RFC 3161 specifies trusted timestamp tokens, while Certificate Transparency illustrates a publicly auditable, append-only Merkle-tree log for TLS certificates \cite{rfc3161,rfc9162}. Project documentation for OpenTimestamps and Chainpoint describes practical hash-anchoring systems in which a data digest is linked to a blockchain-backed proof of prior existence \cite{opentimestamps,chainpoint}. Other anchoring mechanisms, such as trusted timestamping services, transparency logs, append-only audit logs, or permissioned ledgers, may provide similar properties under different trust assumptions. This paper focuses on blockchain anchoring as one deployment assumption: the anchor is operated outside the agent-runtime trust domain and offers independently verifiable inclusion evidence. The same evidence model could also be instantiated with trusted timestamping, transparency logs, append-only audit logs, or permissioned ledgers, provided that their trust, availability, ordering, and governance assumptions are made explicit. An append-only audit log operated within the same trust domain may provide tamper evidence but does not, by itself, remove the operator from the trust model. Recent arXiv preprints have applied related ideas to agentic and multi-agent systems. Jan et al. propose a LangChain-based multi-agent architecture with permissioned blockchain monitoring for policy enforcement and tamper-evident auditability of perception-reasoning-action pipelines \cite{jan2025blockchainagentic}. Li proposes TrustTrack, a protocol for verifiable multi-agent systems that embeds agent identity, policy commitments, and tamper-resistant behavioral traces into agent infrastructure \cite{li2025trustnative}. In contrast, this paper does not define a specific agent framework, product, or ledger. Its contribution is a product- and vendor-independent evidence architecture for GRC-relevant agentic processes, with emphasis on human approvals, agent-to-agent communication, tool-mediated actions, temporal ordering, and incident reconstruction.

We assume an organization operates one or more AI agents with access to internal tools, repositories, communication channels, documents, or APIs. Relevant threats include alteration or deletion of logs after an incident, incomplete capture of agent-to-agent communication, unauthorized communication channels, spoofed or manipulated tool-call records, unclear human-in-the-loop approvals, mismatch between an approved action and the executed action, and loss or modification of intermediate artifacts needed for incident reconstruction.

The proposed evidence layer does not prevent these events. Prevention still depends on secure architecture, access control, sandboxing, monitoring, policy enforcement, and safe tool design. The narrower objective is to make selected critical records tamper-evident so that later audits and investigations can distinguish an unchanged artifact from one that no longer matches its committed state.

\section{Agentic Evidence Architecture}
We propose the pipeline
\begin{center}
\textbf{Capture $\rightarrow$ Canonicalize $\rightarrow$ Hash $\rightarrow$ Anchor $\rightarrow$ Certify $\rightarrow$ Verify.}
\end{center}
A selected event is first captured and converted into a stable canonical representation, for example normalized JSON. The canonical artifact is hashed, and the resulting digest or aggregate commitment is anchored in an external evidence layer. A certificate or verification record binds the commitment to metadata such as event type, timestamp, system context, and verification method. Later, an auditor or incident responder can recompute the hash of the retained artifact and compare it with the earlier commitment.

The canonicalization step is security-relevant rather than merely technical. Agentic events may be represented as JSON objects, log records, transcripts, screenshots, tool-call payloads, or exported audit artifacts. If multiple byte representations can correspond to the same intended semantic event, later integrity verification becomes ambiguous: two systems may hash different byte strings while believing that they refer to the same event. For JSON-based evidence, the JSON Canonicalization Scheme provides an example of a deterministic serialization approach designed to create a stable byte representation for cryptographic operations \cite{rfc8785}. Other evidence types require equivalent normative representations, including explicit rules for ordering, character encoding, timestamps, metadata fields, redaction, and attachment handling before hashing and anchoring.

A minimal event object may contain:\footnote{SHA-512 is used illustratively as a conservative digest choice for long-lived records. This is a design assumption rather than a claim that SHA-256 is currently unsuitable. NIST continues to approve the SHA-2 family, including SHA-256 and SHA-512, for secure-hash applications. The security effect of quantum search differs between preimage and collision resistance and should be assessed against the intended security lifetime. \cite{nistHashPolicy,nist2022postquantum,nistPQCFAQ}}
\begin{verbatim}
{
  "event_id": "ae-0001",
  "event_type": "agent_to_agent",
  "actor": "risk-agent",
  "recipient": "remediation-agent",
  "policy_id": "AI-GOV-07",
  "control_id": "HITL-APPROVAL",
  "artifact_hash": "sha512:<digest>",
  "context_hash": "sha512:<digest>",
  "approval_ref": null,
  "tool_ref": null
}
\end{verbatim}

Three interaction classes are especially relevant. \textbf{Agent--agent} evidence covers messages, delegated tasks, intermediate results, and handovers. \textbf{Agent--human} evidence covers approval requests, escalations, warnings, and human decisions. \textbf{Agent--system} evidence covers tool invocations, API calls, file changes, ticket updates, and execution results. For high-volume environments, individual events can be aggregated into Merkle trees and only the Merkle root anchored, retaining per-event inclusion proofs off-chain.

A critical design principle is data minimization. Sensitive prompts, personal data, trade secrets, and raw transcripts should generally remain in controlled enterprise storage. The public or external evidence layer should contain only hashes or other cryptographic commitments. The retained off-chain artifact and access-control model remain necessary for meaningful verification.

\paragraph{Authenticity and authorized anchoring.}
Anchoring a hash proves integrity of the committed artifact, but not by itself the authenticity of the capture process. A compromised runtime, logging pipeline, or evidence controller could suppress the actual communication and anchor a benign substitute record. The architecture must therefore distinguish artifact integrity from source authenticity and capture authenticity. Evidence events should be bound to authenticated agent, human, or system identities before hashing, and anchoring transactions should be attributable to authorized evidence components. Practical controls may include signed event records, hardware-backed keys, role-based anchoring permissions, separation of duties, append-only local buffers, and independent monitoring of the evidence pipeline itself. Remote attestation architectures provide a model for generating and appraising evidence about the state of a component. Threshold-signature protocols such as FROST may reduce single-key compromise risk, but suitability depends on key governance, participant independence, recovery procedures, and deployment assumptions \cite{rfc9334,rfc9591}.

\section{Temporal Ordering}
Blockchain anchoring can provide an external temporal reference for selected evidence artifacts. The relevant anchor time should be defined as the time at which the commitment obtains verifiable inclusion and, where applicable, finality in the selected anchoring system. The committed digest existed no later than that anchor time. This is not necessarily the exact creation time of the underlying agentic event, and it does not by itself prove semantic causation between messages or actions.

In concurrent agentic systems, multiple agents may submit anchoring transactions in parallel. The resulting on-chain order can reflect transaction propagation, batching, fee selection, confirmation latency, or availability of the anchoring service rather than the semantic order of the underlying workflow. Therefore, blockchain anchoring provides temporal evidence, but causal traceability requires an explicit architectural ordering model.

Direct agent-to-chain anchoring may be sufficient for existence and integrity proofs. Stricter process reconstruction, however, should use an evidence controller or audit gateway that assigns sequence numbers, applies policy checks, links events by hash, and anchors either individual commitments or Merkle-batched aggregates. A chained event can explicitly reference a prior event:
\begin{verbatim}
{
  "event_id": "ae-0002",
  "sequence_number": 42,
  "previous_event_hash": "sha512:<digest>",
  "message_hash": "sha512:<digest>",
  "anchor_ref": "tx:<ref>"
}
\end{verbatim}
Such chaining still does not prove causality in a philosophical or legal sense. It does, however, support causal traceability by showing temporal order, declared dependency, and integrity of the preserved communication and action records. The blockchain can anchor evidence in time. The architecture must define how events are ordered.

\section{GRC and Compliance Verification}
From a Governance, Risk, and Compliance perspective, agentic systems require not only runtime controls but also \emph{evidence-producing controls}. Conventional logs indicate what the system recorded. An independently anchored commitment can additionally demonstrate that a retained record has not changed since a defined point in time.

For \textbf{governance}, this enables critical actions to be linked to roles, policy versions, approval requirements, and decision contexts. For \textbf{risk management}, it supports forensic reconstruction of suspicious agent behavior and allows investigators to compare the final record with evidence committed during execution. For \textbf{compliance}, evidence events can be mapped to control objectives: Was a human approval obtained before a high-impact action? Did a review agent act before an execution agent? Do the executed tool parameters match the approved action? Was an exception or escalation recorded before bypassing a control?

\begin{table}[H]
\centering
\setlength{\abovecaptionskip}{2pt}
\setlength{\belowcaptionskip}{0pt}
\scriptsize
\begin{tabularx}{\columnwidth}{@{}>{\raggedright\arraybackslash}p{0.24\columnwidth}>{\raggedright\arraybackslash}X>{\raggedright\arraybackslash}p{0.30\columnwidth}@{}}
\toprule
\textbf{Control objective} & \textbf{Event to preserve} & \textbf{Evidence} \\
\midrule
Human oversight & Approval request and decision & Hash of request/approval \\
Policy conformance & Required workflow step & Policy/control ID plus event hash \\
Tool control & External action & Parameters/result commitments \\
Segregation of duties & Review before execution & Ordered handover proofs \\
Incident analysis & Suspicious activity & Tamper-evident event timeline \\
Audit readiness & Evidence request & Verifiable certificates/proofs \\
\bottomrule
\end{tabularx}
\caption{Illustrative GRC mappings for agentic evidence.}
\label{tab:grc}
\end{table}

This evidence does not itself establish legal or policy compliance: a control can be faithfully documented and still be incorrectly designed or applied. Its value is evidentiary. It enables an auditor to test whether a required event occurred in the documented sequence and whether the supporting artifact remained unchanged.

\paragraph{Risk-based evidence selection.}
A black-box architecture should not treat every agentic event as equally relevant. Anchoring every message, intermediate state, or low-risk interaction may create unnecessary cost, noise, and privacy exposure. Evidence generation should therefore follow a risk-based policy. Routine agent communication may be retained locally and included in periodic Merkle batches, while high-impact events such as external tool invocations, privilege changes, human approvals, policy exceptions, data transfers, or incident indicators may require immediate or individual anchoring. This aligns the evidence layer with GRC objectives: tamper-evident proofs focus on events relevant for auditability, compliance testing, incident reconstruction, and accountability, while avoiding indiscriminate recording of sensitive or low-value information.

\paragraph{Monitoring evidence stream.}
The verification process can itself produce evidence. A monitoring component may periodically verify whether retained messages, tool-call records, approvals, and process artifacts still match their anchored commitments, whether sequence links are complete, and whether expected policy-relevant events were captured, where an independent source defines the expected event set. The resulting verification reports can be hashed and anchored as a separate monitoring evidence stream. This creates a tamper-evident record not only of agentic communication, but also of the supervisory process that evaluated it. Such secondary evidence does not prove semantic correctness. It documents when and how integrity, completeness, ordering, and policy-conformance of selected records were assessed.

\section{Regulatory Reporting and Evidence Readiness}
The same evidence layer can support regulatory incident workflows. The regulatory regimes differ in scope, thresholds, addressees, and terminology, but they share an operational need for reliable records concerning detection, chronology, impact, decisions, and remediation.

The EU AI Act requires high-risk AI systems to technically allow automatic recording of events (logs) over the system lifetime and imposes record-retention obligations in relevant cases \cite{euai2024}. It also establishes serious-incident reporting duties for providers of high-risk AI systems under Article 73. Reporting depends on the role, system classification, incident type, and statutory trigger. The regulation provides a general outer deadline of 15 days after awareness, with special timing rules for certain severe incidents \cite{euai2024}. For agentic systems within scope, cryptographic commitments to selected logs, approvals, and system states could strengthen the evidentiary basis for a later serious-incident assessment or investigation.

NIS2 requires essential and important entities to notify significant incidents through a staged process: an early warning within 24 hours, an incident notification within 72 hours, and, generally, a final report within one month \cite{nis22022}. As a directive, its concrete operational implementation depends on national transposition and competent authorities. A preserved, tamper-evident agentic timeline can help establish when the organization detected an incident, what evidence was available at each reporting stage, which automated or human decisions were taken, and how remediation evolved.

The CRA creates notification duties for manufacturers concerning actively exploited vulnerabilities and severe incidents affecting the security of products with digital elements. Article 14 applies from 11 September 2026, while the Regulation generally applies from 11 December 2027. The CRA introduces staged reporting obligations, including an early warning within 24 hours and a notification within 72 hours of the manufacturer's awareness. Final-report timing depends on the type of event and report \cite{cra2024,craReporting2026}. Where AI agents participate in vulnerability handling, product-security triage, remediation, or disclosure workflows, anchored evidence can support later verification of detection, triage, mitigation, and communications.

Accordingly, the proposed architecture should not be characterized as a compliance engine. A more precise role is an \emph{evidence layer for compliance-relevant processes}. It may support regulatory reporting readiness where the organization is otherwise subject to the relevant legal regime. It does not determine whether an incident is reportable, identify the responsible operator, or replace the applicable reporting procedure. Its value is to preserve proofs that a record, decision, approval, or tool action existed in a particular form at a particular time.

\section{Illustrative Incident Scenario}
The following scenario is illustrative and does not represent an implemented evaluation. Consider a security-operations workflow using several agents. A detection agent identifies a suspicious cloud configuration. A risk agent reviews the context and recommends remediation. A human analyst approves the proposal under a policy requiring human authorization. An execution agent then invokes a cloud-management API. Finally, a verification agent checks the resulting state.

The evidence layer can anchor commitments to six artifacts: the initial finding, the agent-to-agent handover, the recommendation and policy context, the human approval, the executed tool-call parameters, and the resulting state. If a later incident investigation reveals that an internal transcript was edited, or if an auditor questions whether human approval preceded execution, the retained artifacts can be re-hashed and compared with their earlier commitments.

This model is directly relevant to the lessons of the Hugging Face incident. The independent investigation highlighted both unsanctioned inter-agent communication and attempts to spoof or alter transcript evidence \cite{metrredwood2026hf}. An external tamper-evident commitment would not have prevented the unauthorized communication, but it could reduce the ability to rewrite the evidentiary history of communications or tool records that were captured and anchored before tampering.

\section{Limitations and Future Work}
Several limitations define the scope of the approach. First, hash anchoring proves integrity and temporal existence, not semantic truth, correctness, or authentic capture. Second, the value of the evidence depends on capture completeness. Events that are never captured, or are replaced before capture, cannot be recovered by the anchoring layer. Third, agent and human identities, authorization, policy enforcement, ordering models, and secure time sources remain surrounding-system concerns. Fourth, confidentiality requires disciplined off-chain storage and data minimization. Fifth, the architecture introduces operational questions around event selection, latency, batching, cost, key management, retention, and regulatory deletion requirements.

Future work should evaluate canonical event schemas, canonicalization test suites, Merkle batching, authenticated capture, trusted evidence controllers, cross-system correlation, monitoring evidence streams, policy-to-evidence mappings, and quantitative overhead in realistic multi-agent environments. Future research should also develop a formal evidence model that distinguishes artifact integrity, temporal existence, provenance, capture completeness, authorization, and semantic validity. Such a model could clarify which assurance claims are supported by hash anchoring, which require additional controls, and which cannot be established cryptographically. A further research question is whether hardware-backed keys, remote attestation, threshold signatures, or independently operated controllers can reduce the risk that a compromised component anchors a valid hash of an inauthentic record. Threshold signatures and independent evidence controllers should be evaluated separately, since multiple key shares can still be governed within the same compromised environment \cite{rfc9334,rfc9591}.

\section{Conclusion}
Agentic AI systems create a need for verifiable process memory. As agents communicate, delegate tasks, request approvals, and invoke tools, organizations must be able to reconstruct critical process traces, assess their temporal order, and determine whether supporting records changed after the fact. Conventional logging remains essential, but an externally anchored integrity layer can add a different assurance property: tamper evidence.

We proposed a black-box architecture for agentic processes: a blockchain-anchored evidence layer for AI agent communication, human-in-the-loop decisions, tool-mediated actions, GRC testing, and incident reconstruction. The central principle is deliberately narrow:
\begin{center}
\textbf{Do not only log agentic processes. Make critical traces verifiable.}
\end{center}

\vspace{-0.2em}
\noindent\fbox{\begin{minipage}{0.96\linewidth}
\fontsize{7.2pt}{7.2pt}\selectfont\textbf{Declaration on the Use of AI Tools.}
AI language models were used as tools during the preparation of this working paper, in particular OpenAI GPT-5.5 and Anthropic Claude Sonnet 5. They supported language drafting, critical review of argumentation, source checking, LaTeX/PDF artifact generation, and discussion of examples. The author remains solely responsible for the content, conceptual decisions, source selection, and final version.
\end{minipage}}

\end{document}